\documentclass[12pt]{article}

\usepackage{graphicx}
\usepackage{graphics}
\usepackage{epsfig}
\usepackage{amssymb}
\usepackage{amsmath}
\usepackage{color}
\usepackage{textcomp}
\usepackage{latexsym}

\usepackage[margin=2.5cm,footskip=0.6cm]{geometry}
\begin{document}

\begin{center}
{\Large{\bf Frictional Cooling of Granular Gases: A Molecular Dynamics Study}} \\
\ \\
\ \\
by \\
Prasenjit Das$^1$, Moshe Schwartz$^{2,3}$ and Sanjay Puri$^1$\\

$^1$School of Physical Sciences, Jawaharlal Nehru University, New Delhi 110067, India. \\
$^2$Beverly and Raymond Sackler School of Physics and Astronomy, Tel Aviv University, Ramat Aviv 69934, Israel.\\
$^3$Faculty of Engineering, Holon Institute of Technology, Golomb 52 Holon 5810201, Holon, Israel.
\end{center}

\begin{abstract}
\noindent We study the free evolution of frictional granular gases using large scale molecular dynamics simulation in three dimensions. The system cools due to solid friction among the interacting particles. At early stages of evolution, the density field remains homogeneous and the velocity field follows the Maxwell-Boltzmann (MB) distribution. However, at later times, the density field shows clustering and the velocity field shows local ordering. The ordering in the velocity field is studied by invoking analogy from phase ordering systems. The equal-time correlation function of velocity field follows dynamical scaling. The correlation length of velocity field, $L_v(t)$, exhibits power law growth: $L_v(t)\sim t^{1/3}$.
\end{abstract}
\newpage
\section{Introduction}\label{sec1}
\textit{Granular materials} are one of the most significant and simple forms of matter~\cite{1}. A granular system consists of a large number of mesoscopic particles of regular and irregular shapes with smooth or rough surfaces. The typical size of a particle in a granular system ranges from few $\mu m$ to few $cm$~\cite{2}. These particles loss their kinetic energy due inelasticity and friction, when they collide and slide past each other, and makes a granular system different from ordinary liquids or gases, where it is assumed that energy remains conserved during the collisions between the atoms and molecules. Also, the typical scale of the energy of a granular system is much higher than the thermal energy, i.e., $k_BT$. Thus, granular systems are athermal in nature, and temperature has no role in the dynamics of granular particles. Due to dissipation and athermal nature, a granular system shows distinct properties from usual solids, liquids, and gases~\cite{3}. For example, a sand pile at rest with a slope less than the angle of repose produces solid like behavior due to static friction. Also, the flow of dry sand through the narrow neck of a sand clock is an example of liquid-like behavior. Low-density granular gases appear in sand storm, dust, and smoke etc. The flow of granular matters is important from the industrial application point of view. For example, pharmaceutical industries rely on the processing of powder and pills. In agricultural industries and food processing industries, seeds, grains, and food stuff are transported and manipulated. Understanding the dynamics of granular matters has importance in studying the early stage evolution of stars, galaxies, and astrophysical dust~\cite{4}. Recently, Brilliantov \textit{et al.} described the size distribution of particles in Saturn's ring using the concept of granular aggregation and fragmentation~\cite{5}.

Theoretically, the problem of granular gases has been extensively studied in past two decades~\cite{6}. In the leading theoretical approach, a granular gas is modeled by a system of inelastic hard spheres, where the collision between a pair of spheres is characterized by a constant coefficient of restitution less than unity~\cite{7,8}. One of the most interesting problem studied within that approach is cooling and pattern formation in density and velocity fields. Particles in the  system loss their kinetic energy due to the inelastic binary collisions. In the early stage of evolution, velocity remains random and density remains uniform. However, system cools due to inelastic collisions, is known as homogeneous cooling state (HCS). Haff's law describes the decay of kinetic energy in HCS~\cite{9}. However, later in time, the density field becomes unstable to fluctuations and the system enters an inhomogeneous cooling state (ICS), where the particle-rich clusters are formed and grow. Temperature of the system shows algebraic decay. Non-unified results are reported in literature for velocity distribution in the ICS, e.g., power laws, stretched exponential, etc., have been reported in various studies~\cite{6}. Puri \textit{et al.} studied the complex pattern formation dynamics of density and velocity fields by invoking analogies from studies of phase ordering dynamics. The growth kinetics of the clusters in density and velocity fields have also been studied~\cite{10,11}. Recently, Schwartz \textit{et al.} presented detailed study of pattern formation in two dimensional granular gases where energy dissipation is modeled by solid friction~\cite{12}.

In the case of dense flows, many particles interact among each other simultaneously and stay in prolonged contact. Thus, the concept of collision is not meaningful in this case. In case of granular gases cooling by inelastic collisions, it is known that evolution of the density, velocity, and granular temperature fields is well described by the macroscopic hydrodynamic equations~\cite{13}. However, finding a macroscopic continuous description of such dense granular systems has been an important and challenging issue in the physics of granular flows. In this paper, we consider the solid friction as the only dissipation mechanism in granular gases. Although we believe that frictional dissipation plays significant role in dense systems, we apply it here to gaseous granular matter, as the only mechanism of mechanical energy dissipation, to study its effect on the cooling properties of such dilute systems. Also, we present results for the phase ordering velocity fields and the corresponding growth dynamics. This work will serve also as a reference point for future work on denser systems.

This paper is organized as follows. In Sec.~\ref{sec2}, we describe details of our molecular dynamics simulation. In Sec.~\ref{sec3}, we present detailed numerical results. Finally in  Sec.~\ref{sec4}, we conclude with a brief summary and discussion.

\section{Details of Molecular Dynamics Simulation}\label{sec2}
In our molecular dynamics model simulation, we consider a system of particles confined in a cubic box of size $L^3$. Particles are spherical in shape, identical in size, and of equal mass $m$. Any two particles with position vectors $\vec r_i$ and $\vec r_j$ interact with via a two-body potential with a hardcore of diameter $R_1$ and a thin shell repulsive potential. To be specific, we consider the following form of interaction potential: 
\begin{align}
	\label{pot}
	V(r) = \left\{
	\begin{array}{lr}
		\infty & : r < R_1\\
		V_0\frac{(r-R_2)^2}{(r-R_1)^2} & :  R_1 \le r < R_2\\
		0 &: r \ge R_2
	\end{array}
	\right.
\end{align}
where $r=\mid\vec r_i - \vec r_j\mid$ is the separation between the two particles, $V_0$ is the amplitude of the 
potential, has the dimension of potential energy and $R_2-R_1 < R_1$, is the typical thickness of the thin repulsive shell. Here, Eq.~(\ref{pot}) is to be taken only as a model of repulsive potential that rises steeply from zero at the outer boundary of the shell to infinity at the hardcore. The normal force applied by particle $i$ to particle $j$ is given by
\begin{eqnarray}
	\label{fn}
	\vec F_{ij}^n(r) = -\vec\triangledown V(r),
\end{eqnarray}
where the gradient is taken with respect to $r_j$. The corresponding solid friction force is given by
\begin{eqnarray}
	\label{ff}
	\vec F_{ij}^f(r) = \mu\mid F_{ij}^n\mid \frac{\vec v_1 - \vec v_2}{\mid \vec v_1 - \vec v_2 \mid},
\end{eqnarray}
where $\vec v_i$ and $\vec v_j$ are the linear velocities of particle $i$ and particle $j$ respectively. Eq.~(\ref{ff}) reduces to well known Coulomb's friction force when the thickness of the repulsive shell tends to zero. In this case, the normal component of relative velocity goes to zero and the frictional force becomes purely tangential to the normal force. For simplicity, we did not consider rotational motion of the grains. We use the following units for various relevant quantities: lengths are expressed in units of $R_1$, temperature in $V_0/k_B$ and time in $\sqrt{mR_1^2/V_0}$. For the sake of convenience and numerical stability, we set $R_1=1$, $R_2=1.1$, $V_0=10$, $k_B=1$ and $m=1$. We implemented the standard velocity Verlet algorithm~\cite{14} to update positions and velocities of MD simulations. The integration time step is $\Delta t=0.0005$. The granular gas consisted of N = 400000 particles confined in a 3D box with periodic boundary conditions. We consider two different friction coefficients $\mu=0.10$ and $0.15$. The volume fraction is $\rho\approx0.269$, which corresponds to number density $\phi\approx0.514$. This means that the box size is $L=92$.

We initialize the system by randomly placing particles in the simulation box, such that there is no overlap 
between the cores of any two particles. We assign equal velocity to all the particles but the velocity vector points in random directions so that $\sum_{i=1}^N \vec v_i = 0$, i.e., we fix the centre of mass velocity equal to zero. The system is allowed to evolve till $t=50$ with $\mu = 0$, i.e., the elastic limit. The system is relaxed to a uniform density state and the velocity follows Maxwell-Boltzmann (MB) velocity distribution, which serves as the initial condition (at $t=0$) for our MD simulation of inelastic spheres with $\mu\ne0$.
\begin{figure}
	\centering
	\includegraphics*[width=0.65\textwidth]{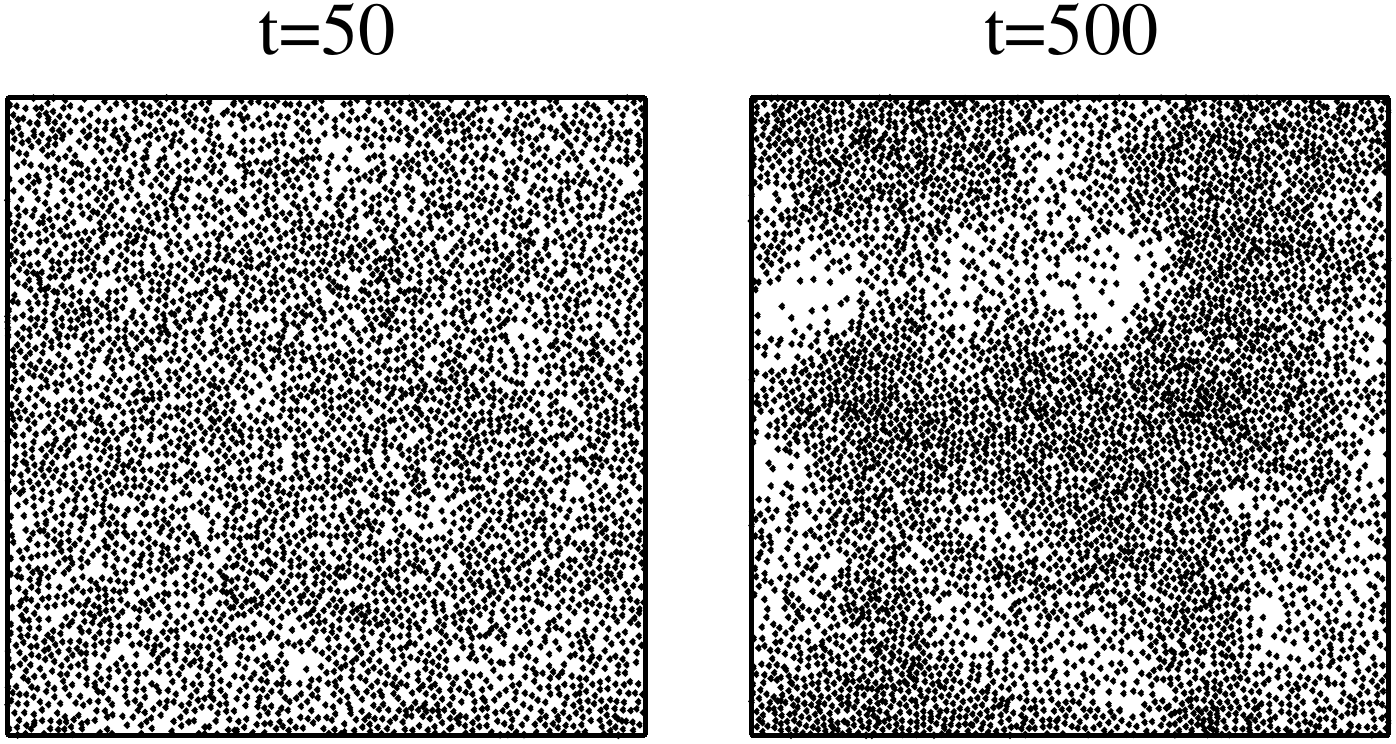}
	\caption{\label{fig1} Two-dimensional (x-y plane) cross section of evolution snapshots of density field at different times, as mentioned. Here, we plot the center of the particles whose z coordinate lies in the range $z\in(31.4,32.6)$. The number of particles in the actual system is $N=400000$ and friction coefficient $\mu=0.10$. The size of the system is $V=92^3$. Void spaces in the density field represents particle free regions.}
\end{figure}

\section{Numerical Results}\label{sec3}
At $t=0$, we start with the homogeneous initial condition for density and velocity fields. Particles dissipates their kinetic energy due to frictional interaction with $\mu\ne0$. In Fig.~\ref{fig1}, we show the evolution snapshots of the density field for $\mu=0.10$. Details are given in the figure caption.

At early times, the density remains roughly homogeneous. However, at later times, the formation of clusters is observed. This cluster formation can be explained as follows. Consider some fluctuations in the homogeneous phase of the density field. Particles in the high-density regions lose more energy than those in the low-density regions. Thus the effective temperature and hence pressure are lesser in the clustered phase than those in the homogeneous phase. This gradient of pressure drives particles from low-density to high-density regions, increasing the density fluctuations. At the late stage, the system becomes sluggish due to the dissipation of energy.

\begin{figure}
	\centering
	\includegraphics*[width=0.65\textwidth]{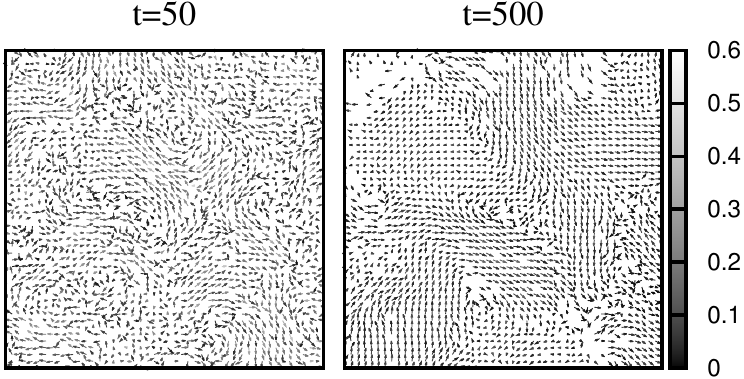}
	\caption{\label{fig2} Evolution snapshots of the coarse-grained velocity field in the x-y plane at different times, as mentioned for $z=32$ layer. The coarse-grained system size is $64^3$. For the sake of clarity, we only plot $48^2$ of $z=32$ layer.}
\end{figure}

At the early stage of evolution, the velocity field remains random. However, at the late stage, local correlation builds up. Due to frictional interaction among the particles, the normal component of relative velocity is dissipated, and the tangential component of relative velocity remains unaltered causing the local parallelization of velocities of two interacting particles. In Fig.~\ref{fig2}, we plot the two dimensional cross-section of the coarse-grained velocity field at different times in the x-y plane.  We divide the system into non-overlapping cubes of size $1.4375^3$ and the average value of velocity is calculated in each cube. The direction of the average velocity is plotted as an arrow starting at the center of each square, using x- and y- components of the velocity. The magnitude of the velocity in each cube is represented by a shade of gray scale. Darker the shade of gray smaller is the velocity. Void spaces in the velocity field are the particle free regions of density field.

\begin{figure}[ht]
	\includegraphics[width=0.50\textwidth]{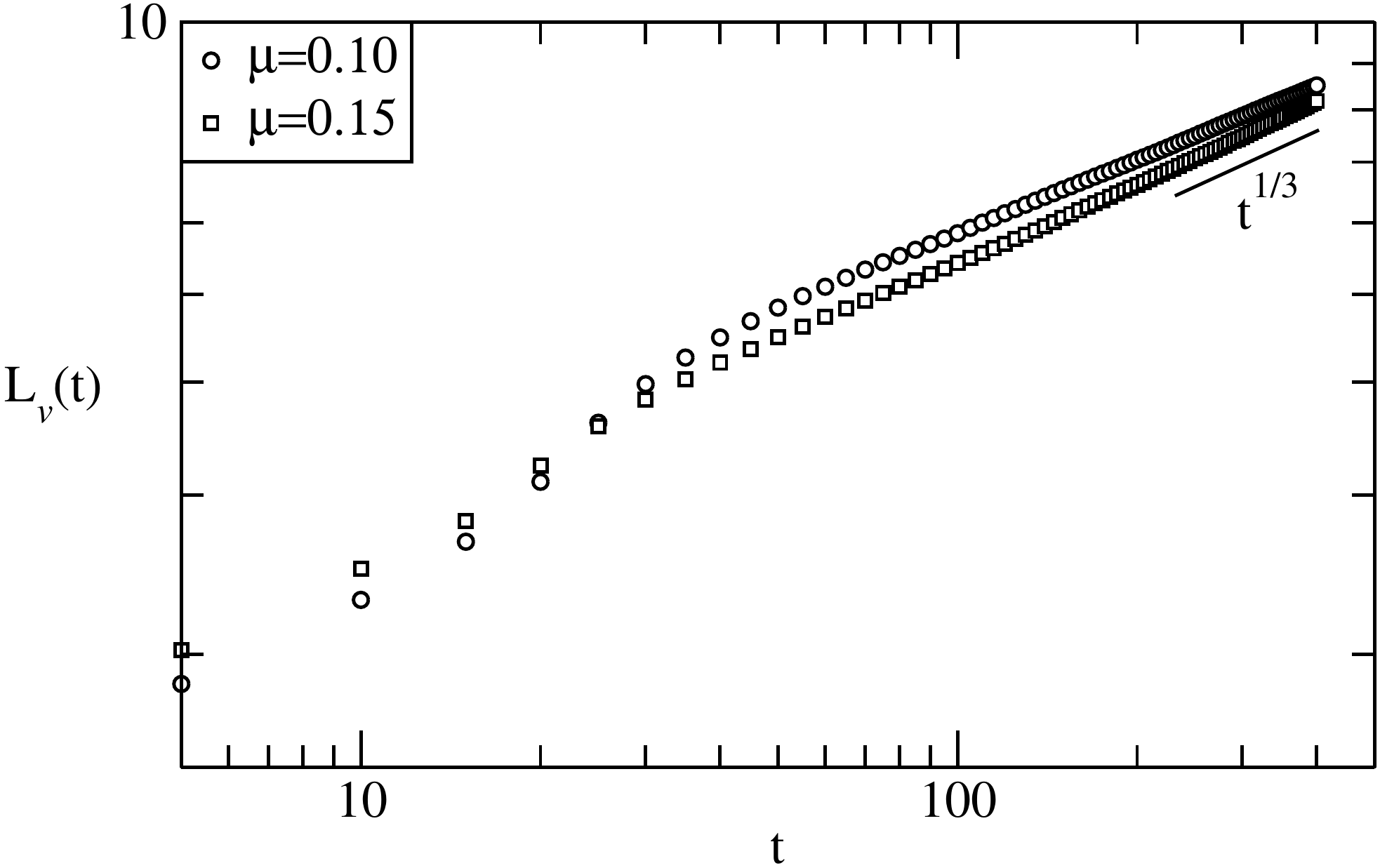}\hspace{1pc}
	\begin{minipage}[b]{18pc}\caption{\label{fig3} Time-dependence of the characteristic length scales $L_v(t)$ of 
			$\vec v(\vec r, t)$-field for $\rho\approx0.269$ and for different $\mu$, as mentioned. Solid line with exponent $\frac{1}{3}$ is shown, corresponds to growth law predicted in the context of plug formation in dense granular flow.}
	\end{minipage}
\end{figure}

We studied ordering in velocity field by invoking analogies from phase ordering kinetics with vector order-parameter field~\cite{15,16,17}. We introduce an order-parameter $\vec v(\vec r, t)$, which is the normalized average velocity in each cell, i.e., we only take the direction of average velocity. This is analogous to the vector order-parameter of dynamical X-Y model in three dimensions~\cite{18}. The morphology of clustering is studied by using equal-time correlation function $C_{vv}(r, t)$, defined as follows
\begin{eqnarray}
	\label{denc}
	C_{vv}(r, t) = \left[\langle\vec v(\vec R, t)\cdot\vec v(\vec R+\vec r, t)\rangle - \langle\vec v(\vec R, t)\rangle\cdot\langle\vec v(\vec R+\vec r, t)\rangle\right],
\end{eqnarray}
where $r$ is the distance between the two points and angular brackets represent the average over different initial conditions~\cite{15,19}. We calculate the characteristic length scale of ordering $L_v(t)$ from $C_{vv}(r, t)$. It is defined as the distance at which correlation decays from 1(at r = 0) to 0.5. The variation of average domain size $L_v(t)$ $vs.$ $t$ for different $\mu$'s are shown in Fig.~\ref{fig3}. Details are given in the figure caption. Line with growth exponent $\alpha=1/3$ in Fig.~\ref{fig3}, corresponds to growth law for plug formation in dense granular flow~\cite{12}.

\begin{figure}
	\centering
	\includegraphics*[width=0.80\textwidth]{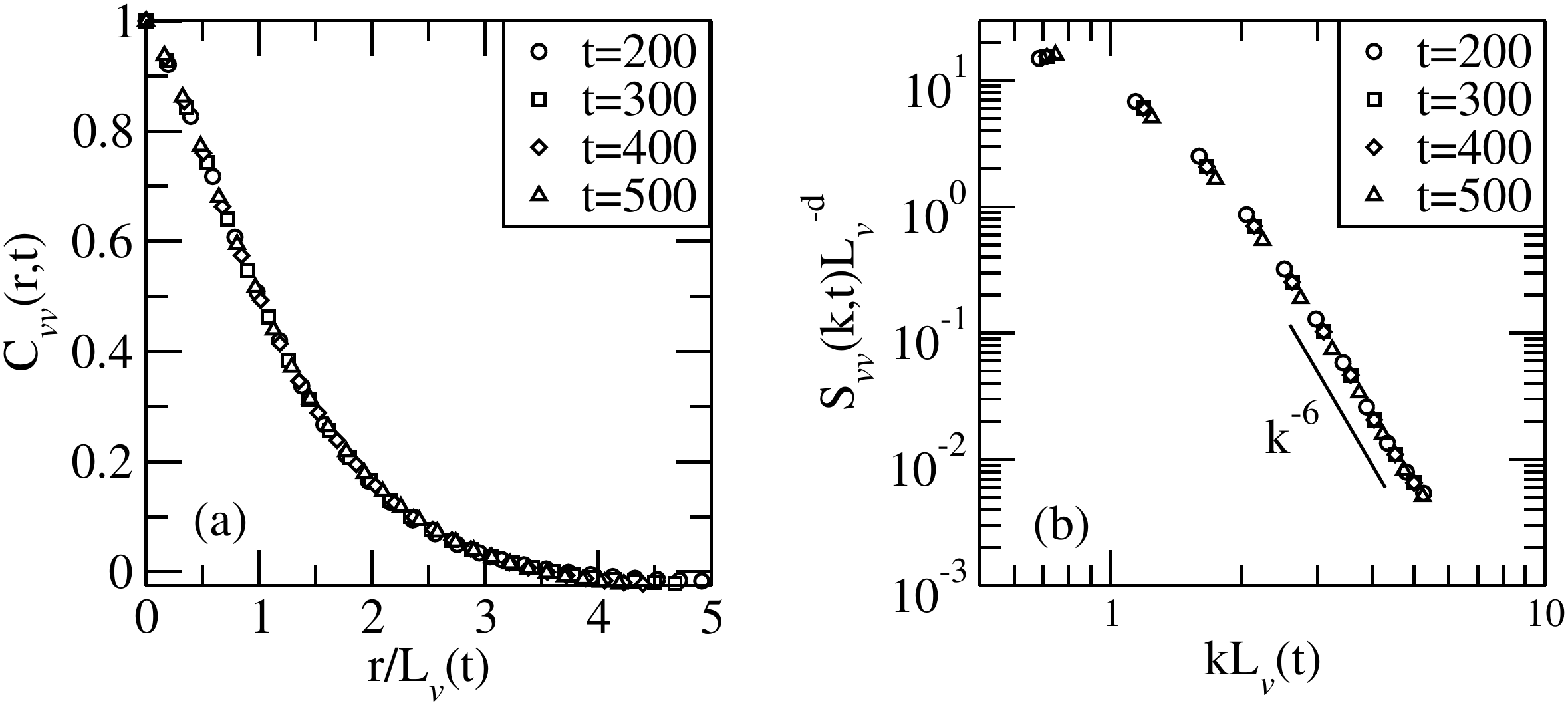}
	\caption{\label{fig4} Scaling plot of spherically averaged correlation functions $C_{vv}(r, t)$ and structure factors $S_{vv}(k, t)$ of coarse-grained $\vec v(\vec r, t)$-field at different times, denoted by the indicated symbols. The coarse-grained system size is $64^3$ and all results are averaged over ten independent runs for $\mu=0.10$ and $\rho\approx0.269$. (a) Plot of $C_{vv}(r, t)$ $vs.$ $r/L_v(t)$ at  different times, as mentioned. Numerical data at different times are indifferent, which confirms dynamical scaling. (b) Plot of $S_{vv}(k, t)L_v^{-d}$ $vs.$ $kL_v(t)$ on log-log scale. Numerical data at different times scaled appropriately, indicate morphological similarity. The line labeled with $k^{-6}$ represents the \textit{generalized Porod's law} with $d=3$ and $n=3$. Rest of the details are same as Fig.~\ref{fig1}.}
\end{figure}

If the clustering is characterized by a single length scale $L_v(t)$, $C_{vv}(r, t)$ shows dynamical scaling given by
\begin{eqnarray}
	\label{dens}
	C_{vv}(r, t) = g_v\left[\frac{r}{L_v(t)}\right],
\end{eqnarray}
where $g_v(x)$ is the master function and $x$ is the time independent scaling variable. Similarly, we calculate \textit{structure factor} $S_{vv}(k, t)$, which is defined as the Fourier transform of the correlation function $C_{vv}(r, t)$ as follows
\begin{eqnarray}
	\label{eqn3}
	S_{vv}\left(\vec k, t\right) = \int d\vec r e^{i\vec k.\vec r}C_{vv}(\vec r, t),
\end{eqnarray}
at wave vector $\vec k$. The corresponding scaling form for $S_{vv}(k, t)$ is given by
\begin{eqnarray}
	\label{eqn4}
	S_{vv}\left(k, t\right) = L_v^d\tilde S_{vv}\left[kL_v(t)\right],
\end{eqnarray}
where $\tilde S_{vv}(p)$ is the scaling function, $d$ is the spatial dimensionality and $p$ is the scaling variable~\cite{15,19}. In Fig.~\ref{fig4}(a), we plot $C_{vv}(r, t)$ $vs.$ $r/L_v(t)$ at different times, as mentioned. We see that numerical data for different times are indistinguishable, which confirms dynamical scaling of ordering. In Fig.~\ref{fig4}(b), we plot $S_{vv}(k, t)L_v^{-d}$ $vs.$ $kL_v(t)$ at different times, denoted by the symbols indicated. In the limit $k\rightarrow\infty$, $S_{vv}(k, t)$ decays as $k^{-6}$, following the \textit{generalized Porod's law} as $S(k, t)\sim k^{-(d+n)}$ with spatial dimensionality $d=3$ and the number of order parameter components $n=3$~\cite{18,20}. This power law decay of $S_{vv}(k, t)$ implies scattering from vortex like defects in the velocity field.

\section{Conclusion}\label{sec4}
Let us conclude this paper with the summary and discussion of our results. We carried out large-scale molecular dynamics (MD) simulation to study the cooling of granular gases, where friction among the interacting particles is considered as the only source of dissipation. At the early stage of evolution, density field remains uniform and velocity field remains random. However, at later times, as the system cools, clustering in density and local ordering in velocity fields have been observed. The equal-time correlation function for velocity field show dynamical scaling, indicating the morphological similarity of ordering. In the limit $k\rightarrow\infty$, structure factor tails for velocity field follows \textit{Generalized Porod's law}: $S_{vv}(k,t)\sim k^{-(d+n)}$, with $d=3$ and $n=3$. The average domain size in velocity field follows power law growth: $L_v(t)\sim t^{1/3}$, proposed in the context of plug formation in dense granular flow. 

\section{Acknowledgments}
PD acknowledges financial support from CSIR, India. The research of MS, grant number 839/14, was supported by the ISF within the ISF-UGC joint research program framework. SP is grateful to UGC, India for support through an Indo-Israeli joint project. He is also grateful to DST, India for support through a J. C. Bose fellowship.

\end{document}